\documentclass[]{spie}  

\usepackage{amsmath,amsfonts,amssymb}
\usepackage{graphicx}
\usepackage[colorlinks=true, allcolors=blue]{hyperref}

\title{The GRANDMA network in preparation for the \\fourth gravitational-wave observing run}

\author[1]{S.~Agayeva}
\author[2,3]{V.~Aivazyan}
\author[4]{S.~Alishov}
\author[5]{M.~Almualla}
\author[6]{C.~Andrade}
\author[7]{S.~Antier}
\author[8]{J.-M.~Bai}
\author[9]{A.~Baransky}
\author[10]{S.~Basa}
\author[11]{P.~Bendjoya}
% ...
\affil[1]{N.Tusi Shamakhy Astrophysical Observatory Azerbaijan National Academy of Sciences, settl.Y. Mammadaliyev, AZ 5626, Shamakhy, Azerbaijan}
\affil[2]{E. Kharadze Georgian National Astrophysical Observatory, Mt.Kanobili, Abastumani, 0301, Adigeni, Georgia}
\affil[3]{Samtskhe-Javakheti  State  University, Rustaveli Str. 113,  Akhaltsikhe, 0080,  Georgia}
\affil[4]{N.Tusi Shamakhy astrophysical Observatory Azerbaijan National Academy of Sciences, settl.Mamedaliyev, AZ 5626, Shamakhy, Azerbaijan}
\affil[5]{American University of Sharjah, Physics Department, PO Box 26666, Sharjah, UAE}
\affil[6]{School of Physics and Astronomy, University of Minnesota, Minneapolis, Minnesota 55455, USA}
\affil[7]{Artemis, Observatoire de la C\^ote d'Azur, Universit\'e C\^ote d'Azur, Boulevard de l'Observatoire, 06304 Nice, France}
\affil[8]{Yunnan Observatories, Chinese Academy of Sciences, Kunming 650011, Yunnan Province, People's Republic of China}
\affil[9]{Astronomical Observatory Taras Shevshenko National University of Kyiv, Observatorna str. 3, Kyiv, 04053, Ukraine}
\affil[10]{Aix Marseille Univ, CNRS, CNES, LAM, IPhU, Marseille, France}
\affil[11]{Laboratoire J.-L. Lagrange, Universit de Nice Sophia-Antipolis, CNRS, Observatoire de la Cote d'Azur, F-06304 Nice, France}
% ...

\authorinfo{Further author information: (Send correspondence to A.d.U.P.)\\A.d.U.P.: E-mail: deugarte@oca.eu}

\begin{document} 
\maketitle

\begin{abstract}
GRANDMA is a world-wide collaboration with the primary scientific goal of studying gravitational-wave sources, discovering their electromagnetic counterparts and characterizing their emission. GRANDMA involves astronomers, astrophysicists, gravitational-wave physicists, and theorists. GRANDMA is now a truly global network of telescopes, with (so far) 30 telescopes in both hemispheres. It incorporates a citizen science programme (Kilonova-Catcher) which constitutes an opportunity to spread the interest in time-domain astronomy. The telescope network is an heterogeneous set of already-existing observing facilities that operate coordinated as a single observatory. Within the network there are wide-field imagers that can observe large areas of the sky to search for optical counterparts, narrow-field instruments that do targeted searches within a predefined list of host-galaxy candidates, and larger telescopes that are devoted to characterization and follow-up of the identified counterparts. Here we present an overview of GRANDMA after the third observing run of the LIGO/VIRGO gravitational-wave observatories in $2019-2020$ and its ongoing preparation for the forthcoming fourth observational campaign (O4). Additionally, we review the potential of GRANDMA for the discovery and follow-up of other types of astronomical transients.
\end{abstract}


\section{INTRODUCTION}
\label{sec:intro}  

Observational techniques in astronomy have greatly evolved since the invention of the telescope, but they have, almost exclusively, involved the study of light. In the last few years, the increased efficiency of gravitational-wave and neutrino observatories are allowing us to use new windows to study the Universe. On some very few occasions we have been able to simultaneously detect individual astronomical sources with several of these \textit{messengers} (photons, gravitational waves, neutrinos), leading to what we now call \textit{multi-messenger astronomy}. Multi-messenger detections are still very rare but provide a wealth of information that can be used to unravel many of the outstanding problems of astrophysics, and are of extreme interest to modern science. To increase our chances of obtaining further multi-messenger data sets, observational astronomers are putting all their efforts into searching for electromagnetic counterparts to gravitational-wave- and neutrino-emitting sources.

Although the sensitivity of neutrino and gravitational-wave observatories are allowing us to routinely detect sources up to cosmological distances, determining a precise localization of the emitting source can be a challenge. This is especially complicated in the case of gravitational waves, where localization can be as rough as tens, or even hundreds of square degrees. The search for the electromagnetic counterparts to these events is further complicated by the fact that they evolve rapidly. In the case of the kilonova emission associated with neutron-star mergers, the peak of the optical emission is reached hours after the GW event and decays rapidly until their light is no longer detectable, by even the largest telescopes, a few days later.

Electromagnetic follow-up observations require the use of well-designed observation techniques. There are two main approaches to this problem: (1) Attempt to image the full uncertainty area of the GW detection as soon and as deep as possible. (2) Use the distance information extracted from the gravitational-wave detection to select candidate galaxies from a catalog and limit the observations to these galaxies. The second approach was successfully used during the O2 observing run of LIGO and VIRGO in 2017 to detect the counterpart of GW170817\cite{Abbott2017_GW170817}, both as a short gammma-ray burst (GRB) at high energies, GRB\,170817A\cite{Abbott2017_GRB170817}, and a kilonova at optical and NIR wavelengths, AT\,2017gfo\cite{Coulter2017Sci}. However, this technique is only efficient for nearby events, at distances lower than $\sim100$ Mpc. At further distances the galaxy catalogs with accurate redshifts begin to be incomplete, and the spatial volume to be covered increases, leading to a steep increment of the number of galaxies that need to be observed and thus making the method less efficient. In O4, where the detection threshold for neutron-star mergers will be close to 200 Mpc, we will have to rely mostly on the first method, covering the full error area of the GW detection with wide-field telescopes.

Instead of investing in developing new dedicated instrumentation, capable of producing the required observations, the GRANDMA (Global Rapid Advanced Network Devoted to the Multi-messenger Addicts) relies on existing telescopes (professional and amateur) around the world that are coordinated to work as a single facility to respond to multi-messenger alerts. A common scheduler designs the observations so that each telescope can contribute to the data collection for each event. Common analysis tools, together with a centralized database to store the data ensure an homogeneous analysis. Finally, our optical observations together with possible gamma-ray public data can be combined in a single multi-physics framework (NMMA) to better understand the astrophysical scenario. In this paper, we will focus on the developments of tools within GRANDMA to prepare for the fourth observing run of gravitational-wave detectors. 

\section{The Telescope network}

The GRANDMA collaboration is a continuously evolving world-wide network that currently includes 30 telescopes within 23 observatories. The scientific team is formed by researchers from 42 institutions, in 18 countries. It was operating during the O3 gravitational-wave observation run of LIGO/VIRGO, when it performed extensive follow-up work of the gravitational-wave alerts\cite{Antier2020a,Antier2020b}. 

Together, the different GRANDMA facilities provide large amounts of observing time that can be allocated for photometric and/or spectroscopic follow-up of transients. The network has access to wide field-of-view telescopes (FoV $>1$ deg$^2$) located on three continents, and remote and robotic telescopes with narrower fields-of-view. 
   \begin{figure} [ht]
   \begin{center}
   \begin{tabular}{c} 
   \includegraphics[width=12cm]{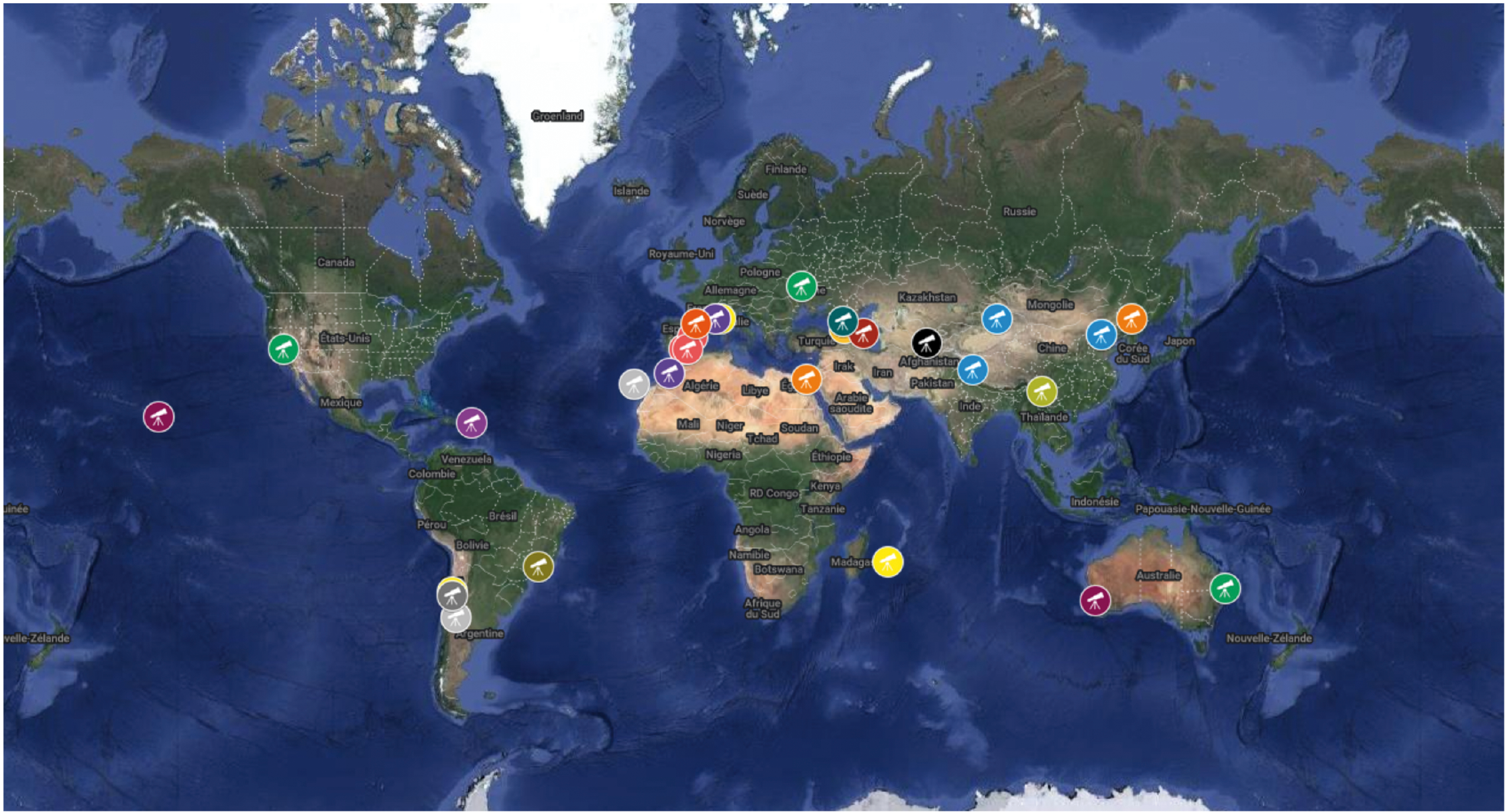}
	\end{tabular}
	\end{center}
   \caption{ \label{fig:GRANDMAworld} 
World map showing the current distribution of GRANDMA observing facilities, not including the Kilonova-Catcher amateur telescopes.}
   \end{figure}

Within GRANDMA, a citizen science programme called \textit{Kilonova-Catcher} (KNC) integrates amateur astronomers spread all over the world that contribute with their observations to the scientific goals of the project. The amateur community is highly motivated and has shown to be capable of meeting the challenge of obtaining data with a quick reaction time and generating high-quality data sets. An increasing number of telescope setups belonging to amateur astronomers or amateur organizations are equipped with digital cameras able to produce deep and reliable observations. KNC has a dedicated portal\footnote{Kilonova Catcher is available at \url{http://kilonovacatcher.in2p3.fr/}} to organize these activities and currently includes around 60 amateur telescopes spread across the globe. After a quality check, their images are treated by GRANDMA in the similar way as images from professional telescopes.

\section{Infrastructure, methods and tools}

The infrastructure for observations with a heterogeneous network requires: i) a central system that coordinates and collects the observations at the network ; ii) management of online processing for the detection per telescope frame and the classification of the electromagnetic (EM) counterpart candidates with rapid diagnostics to enable further decisions for pursuing or interrupting the ongoing observations. The data stored in the infrastructure feeds both, i) and ii), analyses. 

During the O3 observing run, we were using \href{https://gitlab.in2p3.fr/icare/icare}{ICARE} as our central system. In O4, we are developing a completely new system, {\sc SkyPortal}\footnote{{{\sc SkyPortal} is available at \url{https://github.com/skyportal/skyportal}}} \cite{skyportal2019}. {\sc SkyPortal} will distribute GW follow-up observations, process the GW alert information, and also receive the current status of the various observatories in terms of availability and weather forecast. {\sc SkyPortal}  presents a public interface for sharing follow-up observations and associated products and trigger new observations on potential GW counterpart candidates.

\subsection{Alert ingestion and \textit{Fink}}

 {\sc SkyPortal} can process alerts associated with gravitational-wave and $\gamma$-ray burst events. The alerts are received using {\sc PyGCN}\footnote{{{\sc PyGCN}}}. Within GRANDMA, we have also developed the ability to ingest optical alerts from the Zwicky Transient Facility (ZTF), processed by the \textit{Fink} Broker\cite{Moeller2021}. This is the {\sc SkyPortal Fink Client}\footnote{The {{\sc SkyPortal Fink Client} is available at \url{https://github.com/skyportal-contrib/skyportal-fink-client}}}. The advantage is to associate serendipitous kilonovae observed by ZTF and GW events. Alerts from the {\sc SkyPortal Fink Client} contain a unique object identifier, which is used to create a new candidate and a new source, if they do not already exist in the database. If the source already exists, only the new photometric points are added to the existing data set. The \textit{Fink}{} Broker annotates alerts with additional information, such as the result of machine-learning models to classify alerts as kilonovae or supernovae, etc. Combining this with the original data from the ZTF alert, using the \textit{Fink}-filters package, \textit{Fink}'s classifications are processed and added in {\sc SkyPortal}. As more and more data is gathered on a source, the classification provided by \textit{Fink} evolves, and therefore the classification of a source is updated when a new alert is added.

\subsection{Observational strategies}

A centralized system receives the gravitational-wave alerts (or other alerts) and processes an automated joint observation plan using various strategies for efficient scheduling. The first strategy for observations uses wide field-of-view telescopes that blindly scan the search area. The second strategy uses galaxy-targeted follow-up using galaxy catalogs, such as MANGROVE \cite{mangrove}. For both strategies, we use the \href{https://github.com/mcoughlin/gwemopt}{gwemopt} open-source software package for maximizing the probability for joint detection of KN, GRB and GW emission, and combining tiling and time-allocation schemes\cite{CoTo2018,CoAn2019}. The software takes into account the characteristics of each telescope (i.e., field-of-view and image-depth capabilities), visibility of the target, etc. A shortcoming of the current observation scheduling paradigm is the lack of feedback into the schedule from ongoing observations, which means that it may schedule tiles that were already observed rather than prioritizing unobserved tiles. That is why in our new approach for O4, we recover images and their associated products taken by optical surveys such as ATLAS and ZTF, as well as integrating the previous or on-going GRANDMA observations prior to observations. This scheduling software has been developed within {\sc SkyPortal}.

\subsection{Images}

In the past, GRANDMA operations during a campaign have taken advantage of tools such as {\sc Slack} for communication within the team, and on {\sc ownCloud} to store data from observations. Nowadays, both the image storage and chat are encapsulated within {\sc ICARE}, the specific version for GRANDMA of {\sc SkyPortal}. We extract from the data cutout images containing interesting sources, and photometric measurements. We also record the list of observations that have been performed, and group the different optical transients found in a time window (e.g., from the GW event to 10 days) and in a localization area (e.g., the GW sky localization area) to find associations.

\subsection{Optical Data analysis}
\label{sect:analysis}

To uniformly process the diverse set of images acquired by various telescopes, we have developed two dedicated data analysis tools: {\sc MUphoten}\footnote{{{\sc MUphoten} is available at \url{https://gitlab.in2p3.fr/icare/MUPHOTEN}}} \cite{muphoten,Aivazyan2022} and {\sc STDPipe}\footnote{{{\sc STDPipe} is available at \url{https://github.com/karpov-sv/stdpipe}}}\cite{stdpipe,Aivazyan2022}. They follow slightly different approaches, with the former being a ready-to-use set of scripts pre-configured for processing the data from selected instruments, while the latter is a library of both low- and high-level routines for quick creation of custom pipelines for the data from arbitrary telescopes and varying complexity of the analysis (e.g., taking into account spatial dependence of photometric zero points or color terms, using custom noise models, advanced filtering of detected transient candidates, etc.). We foresee them being used by both telescope teams lacking their in-house pipelines for quick-look data analysis, and for the final centralized data processing.

Both pipelines expect the data to be pre-processed by an instrument-specific code to perform bias, dark subtraction, and flat-fielding in advance. This step is done on the telescope side prior to uploading the frames to the GRANDMA centralized data storage. Upon uploading there, the images are processed (manually for now, but a dedicated web-based tool for semi-automatic processing of the images and inspecting the results is planned) by one or both pipelines, and processing results are optionally injected into the database to be used to plan the follow-up observations.

   \begin{figure} [ht]
   \begin{center}
   \begin{tabular}{c} 
   \includegraphics[width=16cm]{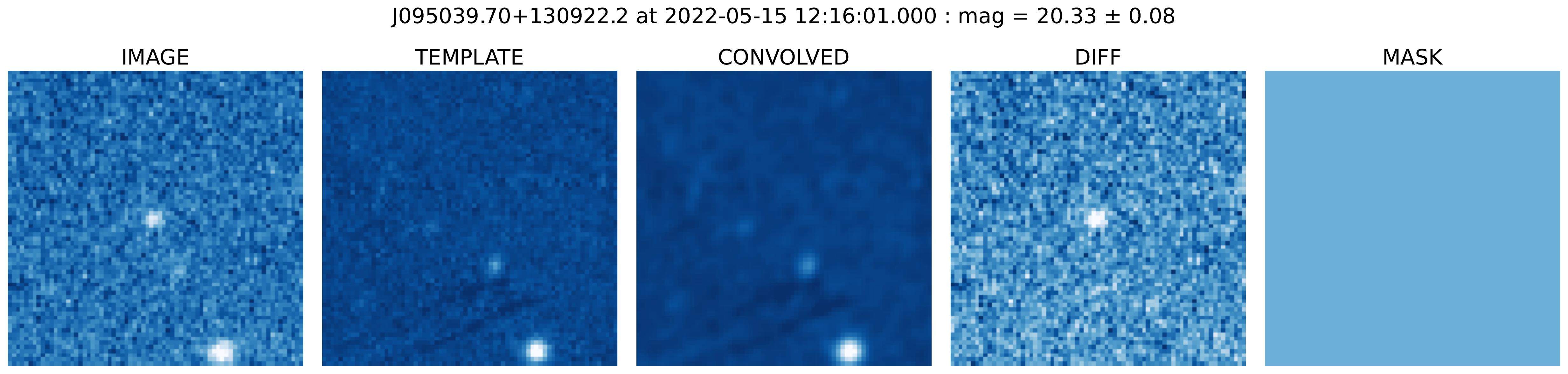}
	\end{tabular}
	\end{center}
   \caption{ \label{fig:220514A} 
    Example of a transient object (GRB\,220514A in this case) as detected and visualized by the {\sc STDPipe} pipeline. The panels show aligned cutouts from the original image, the Pan-STARRS template, the template convolved to match the point-spread function (PSF) of the original image, the difference image, and the mask, all centered on the transient position.}
   \end{figure}

\subsection{Rapid Classification}

 We have also developed online ranking scores to determine the nature of detected transients, and to engage further adequate follow-up. All our scores are running in  {\sc SkyPortal} and imported via \textit{Fink}. The data collected from various telescopes are stored in the central system and used to generate a set of light curves for each transient. We perform linear fits in magnitude versus time/space when light curves have multiple detections over at least a 0.5\,day baseline in a given band. To be as simple as possible, the fits are not weighted and no $\chi$-squared metric is evaluated. We place a hard constraint on fading of at least 0.3\,mag\,day$^{-1}$ in any one of $g^\prime$-, $r^\prime$- or $i^\prime$-bands, as shown to be appropriate for a wide range of kilonova models
 \cite{AnKo2020, KaAn2020}.

\subsection{Human supervision}

During a campaign, GRANDMA needs to contact telescope teams to keep track of what happens and make sure that all the observations run smoothly. For instance, we need to ensure that the observations have been triggered, annotate the transients, adjust their classification and, if needed, request further observations. A team of \textit{follow-up advocates} (team members assuming duties of supervision during operations) coordinate so that there is always at least one person on duty to supervise the operations of the network. To do this, periods of 24 hours are divided in several shifts of the same duration. Each week has a designated shift coordinator, managing groups of follow-up advocates, one for each daily shift. A group of follow-up advocates will be attributed the same shift every day for a week, after which the weekly coordinator and shifters will rotate with another team. A page dedicated to the follow-up advocates has been added to  {\sc SkyPortal}, to manage them directly from the platform using a calendar. 
 {\sc SkyPortal} contains a \textit{notification framework}, accessible to any of its users, that allows them to receive notifications based on certain criteria, using a variety of channels of communication (email, Slack, and/or SMS).
During a campaign, shifters would activate notifications for new events of the notice types that GRANDMA is interested in to ensure that a new event is not missed by any of them.%

\subsection{Interpreting multi-messenger observations and light-curve predictions}

We are currently extending our Bayesian inference nuclear-physics and multi-messenger astrophysics framework, NMMA~\cite{2022arXiv220508513P}. This framework allows to jointly analyze GW, kilonova, and GRB afterglow observations~\cite{Dietrich:2020efo} and even to combine this with nuclear-physics experiments~\cite{Huth:2021bsp}. 
In preparation for O4 and beyond, we further extend the framework and its ingredients to reduce uncertainties in the GW models, the kilonova models, and the description of the GRB afterglow. 

In addition to the possibility to interpret multi-messenger events, the framework can also be used to predict possible electromagnetic counterparts based on low-latency GW information. In particular, based on our knowledge of the equation-of-state of neutron stars, we can make estimates of the masses to predict through phenomenological relations, derived from numerical-relativity simulations, the amount of ejected material and the mass of the possible debris disk formed during the merger process. This information about the material can be related to the expected EM emission~\cite{Stachie:2021noh}.

\section{O4 Preparation campaigns}

In the almost three years between the O3 and the O4 run, the GRANDMA collaboration has been preparing for O4, by updating the scheduling and analysis tools, incorporating new partners, and planning future instrumentation. To test the operations the team has executed two preparation campaigns, with the goal of training and ensure that the observatories, teams and tools are ready for smooth operations during O4. The first campaign was aimed at the follow-up of transient reported by the ZTF survey. The second at the rapid follow-up of GRBs. The results of these campaigns are presented in separate papers\cite{Aivazyan2022}. In this section we discuss some details of the campaigns and the lessons learned from them.

\subsection{Observations of ZTF/FINK transients during Summer 2021}

During a period of six months, between 1 April to 30 September 2021, the GRANDMA network of telescopes produced coordinated observations of transients detected by the Zwicky Transient Facility (ZTF). This work showed the response capability of GRANDMA not only through the use of professional observatories but also amateur ones. In total, 37 observatories were used during this campaign, including both professional (11) and amateur (26) facilities.

Let's recall that \textit{Fink}\cite{Moeller2021} is a community broker designed to filter large time-domain alert streams, such as the current one from the ZTF survey and in the future from the Vera Rubin Observatory. During our campaign, 35 million sources were processed by the broker. \textit{Fink} deals with a large number of topics in the transient sky at all scales, from the Solar System to extragalactic sources. In our case our aim was to attempt observations of kilonova-like events. To this end we use different filters that delivered different numbers of alerts during the campaign: 

\begin{itemize}
    \item Machine learning filter (KN-LC): Delivered 107 candidates.
    \item Nearby Galaxy catalogues-based filter (KN-Mangrove): Delivered 68 candidates.
\end{itemize}

\begin{figure} [ht]
   \begin{center}
   \begin{tabular}{c} 
   \includegraphics[width=12cm]{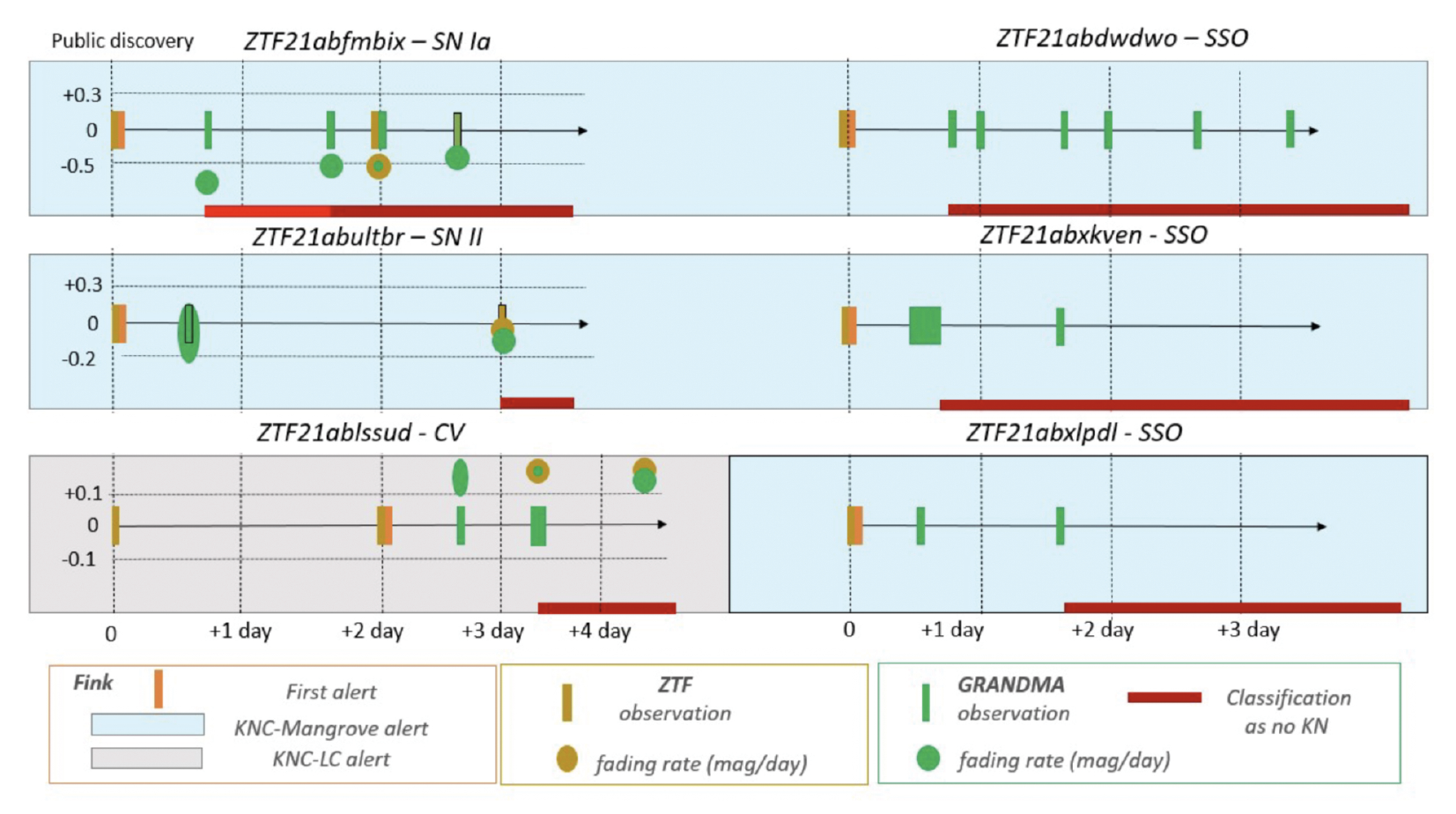}
	\end{tabular}
	\end{center}
   \caption{ \label{fig:fink} 
Overview of the observations of six ZTF alerts distributed to the GRANDMA network after being filtered by the \textit{Fink} broker. The figure, adapted from ref. \cite{Aivazyan2022} displays the times between the distribution of the first alerts and the GRANDMA observations, and how the fading rate of the sources could be calculated, in most cases before the second ZTF visit.}
   \end{figure}

More details on the triggering criteria can be found in the paper that describes the campaign\cite{Aivazyan2022}. Only a handful of candidates were selected by more than one filter. The campaign followed six events for which a total of 180 photometric data points were obtained. The data were uploaded within the first two days and allowed us to determine the decay slopes and complement the sampling of the routine observations obtained by ZTF. The targets were finally classified as Solar System objects, cataclysmic variables, or supernovae. Figure~\ref{fig:fink} summarises the early response of the GRANDMA observations. The campaign showed that, if we respond rapidly and run the reduction pipeline in near real-time, we will be able to filter the most probable candidates before the second observing night. The campaign also demonstrated the ability of amateur astronomers to reach, several times, $>20.5$ mag as image depth, one advantage for the O4 campaign being to be able to search with amateur telescopes for kilonovae located at $\sim180$ Mpc.

\subsection{Gamma-ray burst follow-up during Spring 2022}

The second O4 preparation campaign was aimed at training the team to perform rapid observations. To this end, we decided to follow-up GRB alerts, for which satellites such as the \textit{Neil Gehrels Swift Observatory} generate alerts within dozens of seconds after the event. These sources can be bright during the first minutes after the explosion but decay very rapidly, being excellent targets for training to obtain rapid observations.

The campaign lasted for nine weeks between 20 March and 15 May 2022, a period in which there were eleven GRB alerts. GRANDMA followed eight of these events, those that were observable during night time from the GRANDMA partner observatories. Since this was a training campaign aimed at operations and not at discovery, we did not impose any further filtering of the targets. This meant that a number of the events were found behind strong Galactic extinction and were unlikely to have a detectable counterpart. However, among these eights, we imaged the afterglow of three GRBs within minutes up to hours. All details can be found in our dedicated publication\cite{GRANDMAGRBarticle}.

\section{Expectations for O4 and beyond}

The latest training campaigns have shown that the GRANDMA team is prepared to produce rapid and consistent follow-up observations of astronomical transients. The GRANDMA team is currently working on the development of new and improved capabilities with {\sc SkyPortal} to make observations as user-friendly, and as automatized as possible. Updated versions of {\sc MUphoten} and {\sc STDpipe} are being tested with the existing data of all the GRANDMA facilities to ensure that rapid and efficient photometry can be produced in near real-time during the O4 run. We are continuously making an effort so that the GRANDMA network continues to expand, both with professional and with amateur telescopes. We are aiming at increasing the uniformity of the photometric observations through the purchase and use of standardized sets of filters at those observatories that did not have them. We will also expand our capabilities to query optical surveys and include more available data sets in our database and to complement our target of opportunity observations. To further improve our multi-messenger Bayesian inference framework, we have to extend the incorporated GW, kilonova, and GRB afterglow models to reduce the presence of systematic biases and to allow an accurate interpretation of future detections. 

In the months before the start of the O4 run, we expect to have further training campaigns in parallel to further developments of the multi-physics framework NMMA so that the GRANDMA team will arrive at O4 ready for the discovery of electromagnetic counterparts of gravitational-wave sources.

With the eyes set on O5 we are aiming at the addition of dedicated wide-field telescopes of the 0.8~m class and fields-of-view of two square degrees at sites that will improve the overall coverage of GRANDMA, both in terms of longitudinal coverage, field-of-view, and image depth.

\acknowledgments 
AdUP and SA acknowledge financial support from the Côte D'Azur University through a \textit{CSI recherche} grant awarded to the GRANDMA project (PI: S. Antier).

SA acknowledges the financial support of the Programme National Hautes Energies (PNHE). SA acknowledges the financial support of MITI CNRS Sciences participatives. UBAI acknowledges support from the Ministry of Innovative Development through projects FA-Atech-2018-392 and VA-FA-F-2-010. RI acknowledges Shota Rustaveli National Science Foundation (SRNSF) grant No - RF/18-1193. TAROT has been built with the support of the Institut National des Sciences de l'Univers, CNRS, France. 
MP, SK and MM are supported by European Structural and Investment Fund and the Czech Ministry of Education, Youth and Sports (Projects CZ.02.1.01/0.0/0.0/16\_013/0001403, CZ.02.1.01/0.0/0.0/18\_046/0016007 and CZ.02.1.01/0.0/0.0/15\_003/0000437). 
NBO and DM acknowledge financial support from NASA MUREP MIRO award 80NSSC21M0001, NASA EPSCoR award 80NSSC19M0060, and NSF EiR award 1901296. PG acknowledges financial support from NSF EiR award 1901296. 
DAK acknowledges support from Spanish National Research Project RTI2018-098104-J-I00 (GRBPhot).
XW is supported by the National Science Foundation of China (NSFC grants 12033003 and 11633002), the Scholar Program of Beijing Academy of Science and Technology (DZ:BS202002), and the Tencent Xplorer Prize. 
J.Mao is supported by the National Natural Science Foundation of China 11673062 and the Oversea Talent Program of Yunnan Province.
The work of FN is supported by NOIRLab, which is managed by the Association of Universities for Research in Astronomy (AURA) under a cooperative agreement with the National Science Foundation.
The GRANDMA consortium thank the amateur participants to the \textit{kilonova-catcher} program. The \textit{kilonova-catcher} program is supported by the IdEx Universit\'e de Paris, ANR-18-IDEX-0001. This research made use of the cross-match service provided by CDS, Strasbourg. MC acknowledges support from the National Science Foundation with grant numbers PHY-2010970 and OAC-2117997. 
MC and CA acknowledge support from a ``Preparing for Astrophysics with LSST'' grant with grant number KSI-2.
GR acknowledges financial support from the Nederlandse Organisatie voor Wetenschappelijk Onderzoek (NWO) through the Projectruimte and VIDI grants (PI: Nissanke). Thanks to the National Astronomical Research Institute of Thailand (Public Organization), based on observations made with the Thai Robotic Telescope under program ID TRTC08D\_005 and TRTC09A\_002.
S. Leonini thanks M. Conti, P. Rosi, and L. M. Tinjaca Ramirez.
SA thanks Etienne Bertrand and le ``Club des C\'eph\'eides'' for their observations of ZTF21abxkven.
\bibliography{GRANDMA} 

\begin{thebibliography}{10}

\bibitem{Abbott2017_GW170817}
{Abbott}, B.~P., {Abbott}, R., {Abbott}, T.~D., {Acernese}, F., {Ackley}, K.,
  {Adams}, C., {Adams}, T., {Addesso}, P., {Adhikari}, R.~X., {Adya}, V.~B.,
  {Affeldt}, C., {Afrough}, M., {Agarwal}, B., {Agathos}, M., {Agatsuma}, K.,
  {Aggarwal}, N., {Aguiar}, O.~D., {Aiello}, L., {Ain}, A., {Ajith}, P.,
  {Allen}, B., {Allen}, G., {Allocca}, A., {Altin}, P.~A., {Amato}, A.,
  {Ananyeva}, A., {Anderson}, S.~B., {Anderson}, W.~G., {Angelova}, S.~V.,
  {Antier}, S., {Appert}, S., {Arai}, K., {Araya}, M.~C., {Areeda}, J.~S.,
  {Arnaud}, N., {Arun}, K.~G., {Ascenzi}, S., {Ashton}, G., {Ast}, M., {Aston},
  S.~M., {Astone}, P., {Atallah}, D.~V., {Aufmuth}, P., {Aulbert}, C.,
  {AultONeal}, K., {Austin}, C., {Avila-Alvarez}, A., {Babak}, S., {Bacon}, P.,
  {Bader}, M.~K.~M., {Bae}, S., {Bailes}, M., {Baker}, P.~T., {Baldaccini}, F.,
  {Ballardin}, G., {Ballmer}, S.~W., {Banagiri}, S., {Barayoga}, J.~C.,
  {Barclay}, S.~E., {Barish}, B.~C., {Barker}, D., {Barkett}, K., {Barone}, F.,
  {Barr}, B., {Barsotti}, L., {Barsuglia}, M., {Barta}, D., {Barthelmy}, S.~D.,
  {Bartlett}, J., {Bartos}, I., {Bassiri}, R., {Basti}, A., {Batch}, J.~C.,
  {Bawaj}, M., {Bayley}, J.~C., {Bazzan}, M., {B{\'e}csy}, B., {Beer}, C.,
  {Bejger}, M., {Belahcene}, I., {Bell}, A.~S., {Berger}, B.-K., and {et al.},
  ``{GW170817: Observation of Gravitational Waves from a Binary Neutron Star
  Inspiral},'' {\em Phys. Rev. Lett.}~{\bf 119},  161101 (Oct. 2017).

\bibitem{Abbott2017_GRB170817}
{Abbott}, B.~P., {Abbott}, R., {Abbott}, T.~D., {Acernese}, F., {Ackley}, K.,
  {Adams}, C., {Adams}, T., {Addesso}, P., {Adhikari}, R.~X., {Adya}, V.~B.,
  {Affeldt}, C., {Afrough}, M., {Agarwal}, B., {Agathos}, M., {Agatsuma}, K.,
  {Aggarwal}, N., {Aguiar}, O.~D., {Aiello}, L., {Ain}, A., {Ajith}, P.,
  {Allen}, B., {Allen}, G., {Allocca}, A., {Aloy}, M.~A., {Altin}, P.~A.,
  {Amato}, A., {Ananyeva}, A., {Anderson}, S.~B., {Anderson}, W.~G.,
  {Angelova}, S.~V., {Antier}, S., {Appert}, S., {Arai}, K., {Araya}, M.~C.,
  {Areeda}, J.~S., {Arnaud}, N., {Arun}, K.~G., {Ascenzi}, S., {Ashton}, G.,
  {Ast}, M., {Aston}, S.~M., {Astone}, P., {Atallah}, D.~V., {Aufmuth}, P.,
  {Aulbert}, C., {AultONeal}, K., {Austin}, C., {Avila-Alvarez}, A., {Babak},
  S., {Bacon}, P., {Bader}, M.~K.~M., {Bae}, S., {Baker}, P.~T., {Baldaccini},
  F., {Ballardin}, G., {Ballmer}, S.~W., {Banagiri}, S., {Barayoga}, J.~C.,
  {Barclay}, S.~E., {Barish}, B.~C., {Barker}, D., {Barkett}, K., {Barone}, F.,
  {Barr}, B., {Barsotti}, L., {Barsuglia}, M., {Barta}, D., {Bartlett}, J.,
  {Bartos}, I., {Bassiri}, R., {Basti}, A., {Batch}, J.~C., {Bawaj}, M.,
  {Bayley}, J.~C., {Bazzan}, M., {B{\'e}csy}, B., {Beer}, C., {Bejger}, M.,
  {Belahcene}, I., {Bell}, A.~S., {Berger}, B.~K., {Bergmann}, G., {Bero},
  J.~J., and {et al.}, ``{Gravitational Waves and Gamma-Rays from a Binary
  Neutron Star Merger: GW170817 and GRB 170817A},'' {\em Astroph Journ.}~{\bf
  848},  L13 (Oct. 2017).

\bibitem{Coulter2017Sci}
{Coulter}, D.~A., {Foley}, R.~J., {Kilpatrick}, C.~D., {Drout}, M.~R., {Piro},
  A.~L., {Shappee}, B.~J., {Siebert}, M.~R., {Simon}, J.~D., {Ulloa}, N.,
  {Kasen}, D., {Madore}, B.~F., {Murguia-Berthier}, A., {Pan}, Y.~C.,
  {Prochaska}, J.~X., {Ramirez-Ruiz}, E., {Rest}, A., and {Rojas-Bravo}, C.,
  ``{Swope Supernova Survey 2017a (SSS17a), the optical counterpart to a
  gravitational wave source},'' {\em Science}~{\bf 358},  1556--1558 (Dec.
  2017).

\bibitem{Antier2020a}
{Antier}, S., {Agayeva}, S., {Aivazyan}, V., {Alishov}, S., {Arbouch}, E.,
  {Baransky}, A., {Barynova}, K., {Bai}, J.~M., {Basa}, S., {Beradze}, S.,
  {Bertin}, E., {Berthier}, J., {Bla{\v{z}}ek}, M., {Bo{\"e}r}, M.,
  {Burkhonov}, O., {Burrell}, A., {Cailleau}, A., {Chabert}, B., {Chen}, J.~C.,
  {Christensen}, N., {Coleiro}, A., {Cordier}, B., {Corre}, D., {Coughlin},
  M.~W., {Coward}, D., {Crisp}, H., {Delattre}, C., {Dietrich}, T., {Ducoin},
  J.~G., {Duverne}, P.~A., {Marchal-Duval}, G., {Gendre}, B., {Eymar}, L.,
  {Fock-Hang}, P., {Han}, X., {Hello}, P., {Howell}, E.~J., {Inasaridze}, R.,
  {Ismailov}, N., {Kann}, D.~A., {Kapanadze}, G., {Klotz}, A., {Kochiashvili},
  N., {Lachaud}, C., {Leroy}, N., {Le Van Su}, A., {Lin}, W.~L., {Li}, W.~X.,
  {Lognone}, P., {Marron}, R., {Mo}, J., {Moore}, J., {Natsvlishvili}, R.,
  {Noysena}, K., {Perrigault}, S., {Peyrot}, A., {Samadov}, D., {Sadibekova},
  T., {Simon}, A., {Stachie}, C., {Teng}, J.~P., {Thierry}, P., {Th{\"o}ne},
  C.~C., {Tillayev}, Y., {Turpin}, D., {de Ugarte Postigo}, A., {Vachier}, F.,
  {Vardosanidze}, M., {Vasylenko}, V., {Vidadi}, Z., {Wang}, X.~F., {Wang},
  C.~J., {Wei}, J., {Yan}, S.~Y., {Zhang}, J.~C., {Zhang}, J.~J., and {Zhang},
  X.~H., ``{The first six months of the Advanced LIGO's and Advanced Virgo's
  third observing run with GRANDMA},'' {\em Mon. Not. Roy. Astron. Soc.}~{\bf
  492},  3904--3927 (Mar. 2020).

\bibitem{Antier2020b}
{Antier}, S., {Agayeva}, S., {Almualla}, M., {Awiphan}, S., {Baransky}, A.,
  {Barynova}, K., {Beradze}, S., {Bla{\v{z}}ek}, M., {Bo{\"e}r}, M.,
  {Burkhonov}, O., {Christensen}, N., {Coleiro}, A., {Corre}, D., {Coughlin},
  M.~W., {Crisp}, H., {Dietrich}, T., {Ducoin}, J.~G., {Duverne}, P.~A.,
  {Marchal-Duval}, G., {Gendre}, B., {Gokuldass}, P., {Eggenstein}, H.~B.,
  {Eymar}, L., {Hello}, P., {Howell}, E.~J., {Ismailov}, N., {Kann}, D.~A.,
  {Karpov}, S., {Klotz}, A., {Kochiashvili}, N., {Lachaud}, C., {Leroy}, N.,
  {Lin}, W.~L., {Li}, W.~X., {Ma{\v{s}}ek}, M., {Mo}, J., {Menard}, R.,
  {Morris}, D., {Noysena}, K., {Orange}, N.~B., {Prouza}, M., {Rattanamala},
  R., {Sadibekova}, T., {Saint-Gelais}, D., {Serrau}, M., {Simon}, A.,
  {Stachie}, C., {Th{\"o}ne}, C.~C., {Tillayev}, Y., {Turpin}, D., {Postigo},
  A. d.~U., {Vasylenko}, V., {Vidadi}, Z., {Was}, M., {Wang}, X.~F., {Zhang},
  J.~J., {Zhang}, T.~M., and {Zhang}, X.~H., ``{GRANDMA observations of
  advanced LIGO's and advanced Virgo's third observational campaign},'' {\em
  Mon. Not. Roy. Astron. Soc.}~{\bf 497},  5518--5539 (Oct. 2020).

\bibitem{skyportal2019}
van~der Walt, S.~J., Crellin-Quick, A., and Bloom, J.~S., ``{SkyPortal}: An
  astronomical data platform,'' {\em Journal of Open Source Software}~{\bf 4}
  (may 2019).

\bibitem{Moeller2021}
{M{\"o}ller}, A., {Peloton}, J., {Ishida}, E. E.~O., {Arnault}, C., {Bachelet},
  E., {Blaineau}, T., {Boutigny}, D., {Chauhan}, A., {Gangler}, E.,
  {Hernandez}, F., {Hrivnac}, J., {Leoni}, M., {Leroy}, N., {Moniez}, M.,
  {Pateyron}, S., {Ramparison}, A., {Turpin}, D., {Ansari}, R., {Allam}, Tarek,
  J., {Bajat}, A., {Biswas}, B., {Boucaud}, A., {Bregeon}, J., {Campagne},
  J.-E., {Cohen-Tanugi}, J., {Coleiro}, A., {Dornic}, D., {Fouchez}, D.,
  {Godet}, O., {Gris}, P., {Karpov}, S., {Nebot Gomez-Moran}, A., {Neveu}, J.,
  {Plaszczynski}, S., {Savchenko}, V., and {Webb}, N., ``{FINK, a new
  generation of broker for the LSST community},'' {\em Mon. Not. Roy. Astron.
  Soc.}~{\bf 501},  3272--3288 (Mar. 2021).

\bibitem{mangrove}
{Ducoin}, J.~G., {Corre}, D., {Leroy}, N., and {Le Floch}, E., ``{Optimizing
  gravitational waves follow-up using galaxies stellar mass},'' {\em Mon. Not.
  Roy. Astron. Soc.}~{\bf 492},  4768--4779 (Mar. 2020).

\bibitem{CoTo2018}
Coughlin, M.~W., Tao, D., Chan, M.~L., Chatterjee, D., Christensen, N., Ghosh,
  S., Greco, G., Hu, Y., Kapadia, S., Rana, J., Salafia, O.~S., and Stubbs,
  C.~W., ``Optimizing searches for electromagnetic counterparts of
  gravitational wave triggers,'' {\em Mon. Not. Roy. Astron. Soc.}~{\bf
  478}(1),  692--702 (2018).

\bibitem{CoAn2019}
Coughlin, M.~W. et~al., ``{Optimizing multitelescope observations of
  gravitational-wave counterparts},'' {\em Mon. Not. Roy. Astron. Soc.}~{\bf
  489}(4),  5775--5783 (2019).

\bibitem{muphoten}
{Duverne}, P.~A., {Antier}, S., {Basa}, S., {Corre}, D., {Coughlin}, M.~W.,
  {Filippenko}, A.~V., {Klotz}, A., {Hello}, P., and {Zheng}, W., ``{MUPHOTEN :
  a MUlti-band PHOtometry Tool for TElescope Network},'' {\em arXiv e-prints} ,
   arXiv:2201.07565 (Jan. 2022).

\bibitem{Aivazyan2022}
{Aivazyan}, V., {Almualla}, M., {Antier}, S., {Baransky}, A., {Barynova}, K.,
  {Basa}, S., {Bayard}, F., {Beradze}, S., {Berezin}, D., {Blazek}, M.,
  {Boutigny}, D., {Boust}, D., {Broens}, E., {Burkhonov}, O., {Cailleau}, A.,
  {Christensen}, N., {Cejudo}, D., {Coleiro}, A., {Coughlin}, M.~W.,
  {Datashvili}, D., {Dietrich}, T., {Dolon}, F., {Ducoin}, J.~G., {Duverne},
  P.~A., {Marchal-Duval}, G., {Galdies}, C., {Granier}, L., {Godunova}, V.,
  {Gokuldass}, P., {Eggenstein}, H.~B., {Freeberg}, M., {Hello}, P.,
  {Inasaridze}, R., {Ishida}, E.~O., {Jaquiery}, P., {Kann}, D.~A.,
  {Kapanadze}, G., {Karpov}, S., {Kiendrebeogo}, R.~W., {Klotz}, A., {Kneip},
  R., {Kochiashvili}, N., {Kou}, W., {Kugel}, F., {Lachaud}, C., {Leonini}, S.,
  {Leroy}, A., {Leroy}, N., {Le Van Su}, A., {Marchais}, D., {Masek}, M.,
  {Midavaine}, T., {Moller}, A., {Morris}, D., {Natsvlishvili}, R., {Navarete},
  F., {Noysena}, K., {Nissanke}, S., {Noonan}, K., {Orange}, N.~B., {Peloton},
  J., {Popowicz}, A., {Pradier}, T., {Prouza}, M., {Raaijmakers}, G.,
  {Rajabov}, Y., {Richmond}, M., {Romanyuk}, Y., {Rousselot}, L., {Sadibekova},
  T., {Serrau}, M., {Sokoliuk}, O., {Song}, X., {Simon}, A., {Stachie}, C.,
  {Taylor}, A., {Tillayev}, Y., {Turpin}, D., {Vardosanidze}, M., {Vlieghe},
  J., {Tosta e Melo}, I., {Wang}, X.~F., and {Zhu}, J., ``{GRANDMA Observations
  of ZTF/Fink Transients during Summer 2021},'' {\em arXiv e-prints} ,
  arXiv:2202.09766 (Feb. 2022).

\bibitem{stdpipe}
{Karpov}, S., ``{STDPipe: Simple Transient Detection Pipeline}.'' Astrophysics
  Source Code Library, record ascl:2112.006 (Dec. 2021).

\bibitem{AnKo2020}
{Andreoni}, I., {Kool}, E.~C., {Sagu{\'e}s Carracedo}, A., {Kasliwal}, M.~M.,
  {Bulla}, M., {Ahumada}, T., {Coughlin}, M.~W., {Anand}, S., {Sollerman}, J.,
  {Goobar}, A., {Kaplan}, D.~L., {Loveridge}, T.~T., {Karambelkar}, V.,
  {Cooke}, J., {Bagdasaryan}, A., {Bellm}, E.~C., {Cenko}, S.~B., {Cook},
  D.~O., {De}, K., {Dekany}, R., {Delacroix}, A., {Drake}, A., {Duev}, D.~A.,
  {Fremling}, C., {Golkhou}, V.~Z., {Graham}, M.~J., {Hale}, D., {Kulkarni},
  S.~R., {Kupfer}, T., {Laher}, R.~R., {Mahabal}, A.~A., {Masci}, F.~J.,
  {Rusholme}, B., {Smith}, R.~M., {Tzanidakis}, A., {Van Sistine}, A., and
  {Yao}, Y., ``{Constraining the Kilonova Rate with Zwicky Transient Facility
  Searches Independent of Gravitational Wave and Short Gamma-Ray Burst
  Triggers},'' {\em Astroph. Journ.}~{\bf 904},  155 (Dec. 2020).

\bibitem{KaAn2020}
{Kasliwal}, M.~M., {Anand}, S., {Ahumada}, T., {Stein}, R., {Carracedo}, A.~S.,
  {Andreoni}, I., {Coughlin}, M.~W., {Singer}, L.~P., {Kool}, E.~C., {De}, K.,
  {Kumar}, H., {AlMualla}, M., {Yao}, Y., {Bulla}, M., {Dobie}, D., {Reusch},
  S., {Perley}, D.~A., {Cenko}, S.~B., {Bhalerao}, V., {Kaplan}, D.~L.,
  {Sollerman}, J., {Goobar}, A., {Copperwheat}, C.~M., {Bellm}, E.~C.,
  {Anupama}, G.~C., {Corsi}, A., {Nissanke}, S., {Agudo}, I., {Bagdasaryan},
  A., {Barway}, S., {Belicki}, J., {Bloom}, J.~S., {Bolin}, B., {Buckley}, D.
  A.~H., {Burdge}, K.~B., {Burruss}, R., {Caballero-Garc{\'\i}a}, M.~D.,
  {Cannella}, C., {Castro-Tirado}, A.~J., {Cook}, D.~O., {Cooke}, J.,
  {Cunningham}, V., {Dahiwale}, A., {Deshmukh}, K., {Dichiara}, S., {Duev},
  D.~A., {Dutta}, A., {Feeney}, M., {Franckowiak}, A., {Frederick}, S.,
  {Fremling}, C., {Gal-Yam}, A., {Gatkine}, P., {Ghosh}, S., {Goldstein},
  D.~A., {Golkhou}, V.~Z., {Graham}, M.~J., {Graham}, M.~L., {Hankins}, M.~J.,
  {Helou}, G., {Hu}, Y., {Ip}, W.-H., {Jaodand}, A., {Karambelkar}, V., {Kong},
  A. K.~H., {Kowalski}, M., {Khandagale}, M., {Kulkarni}, S.~R., {Kumar}, B.,
  {Laher}, R.~R., {Li}, K.~L., {Mahabal}, A., {Masci}, F.~J., {Miller}, A.~A.,
  {Mogotsi}, M., {Mohite}, S., {Mooley}, K., {Mroz}, P., {Newman}, J.~A.,
  {Ngeow}, C.-C., {Oates}, S.~R., {Patil}, A.~S., {Pandey}, S.~B., {Pavana},
  M., {Pian}, E., {Riddle}, R., {S{\'a}nchez-Ram{\'\i}rez}, R., {Sharma}, Y.,
  {Singh}, A., {Smith}, R., {Soumagnac}, M.~T., {Taggart}, K., {Tan}, H.,
  {Tzanidakis}, A., {Troja}, E., {Valeev}, A.~F., {Walters}, R., {Waratkar},
  G., {Webb}, S., {Yu}, P.-C., {Zhang}, B.-B., {Zhou}, R., and {Zolkower}, J.,
  ``{Kilonova Luminosity Function Constraints Based on Zwicky Transient
  Facility Searches for 13 Neutron Star Merger Triggers during O3},'' {\em
  Astroph. Journ.}~{\bf 905},  145 (Dec. 2020).

\bibitem{2022arXiv220508513P}
{Pang}, P. T.~H., {Dietrich}, T., {Coughlin}, M.~W., {Bulla}, M., {Tews}, I.,
  {Almualla}, M., {Barna}, T., {Kiendrebeogo}, W., {Kunert}, N., {Mansingh},
  G., {Reed}, B., {Sravan}, N., {Toivonen}, A., {Antier}, S., {VandenBerg},
  R.~O., {Heinzel}, J., {Nedora}, V., {Salehi}, P., {Sharma}, R.,
  {Somasundaram}, R., and {Van Den Broeck}, C., ``{NMMA: A nuclear-physics and
  multi-messenger astrophysics framework to analyze binary neutron star
  mergers},'' {\em arXiv e-prints} ,  arXiv:2205.08513 (May 2022).

\bibitem{Dietrich:2020efo}
Dietrich, T., Coughlin, M.~W., Pang, P. T.~H., Bulla, M., Heinzel, J., Issa,
  L., Tews, I., and Antier, S., ``{Multimessenger constraints on the
  neutron-star equation of state and the Hubble constant},'' {\em Science}~{\bf
  370}(6523),  1450--1453 (2020).

\bibitem{Huth:2021bsp}
Huth, S. et~al., ``{Constraining Neutron-Star Matter with Microscopic and
  Macroscopic Collisions},'' {\em Nature}~{\bf 606},  276--280 (2022).

\bibitem{Stachie:2021noh}
Stachie, C. et~al., ``{Predicting electromagnetic counterparts using
  low-latency gravitational-wave data products},'' {\em Mon. Not. Roy. Astron.
  Soc.}~{\bf 505}(3),  4235--4248 (2021).

\bibitem{GRANDMAGRBarticle}
{GRANDMA Collaboration}, ``{GRB campaign with GRANDMA},'' {\em in prep.}
  (2022).

\end{thebibliography}
\bibliographystyle{spiebib} 

\end{document}